\documentclass[a4paper]{jpconf}
\usepackage[cp1250]{inputenc}
\usepackage[T1]{fontenc}
\newcommand{\ee}{\end{equation}}
\newcommand{\be}{\begin{equation}}
\newcommand{\ba}{\begin{array}}
\newcommand{\ea}{\end{array}}
\newcommand{\m}{M_{H^{\pm}}}
\newcommand{\g}{\,\mbox{GeV}}
\newcommand{\la}{\lambda_1}
\newcommand{\lb}{\lambda_2}
\newcommand{\lc}{\lambda_3}

\newcommand{\lczp}{\lambda_{345}}
\newcommand{\rg}{R_{\gamma\gamma}}

\newcommand{\fr}{\frac}
\usepackage[pdftex]{graphicx}
\usepackage{amsmath}
\usepackage{amssymb}
\usepackage{amsthm}
\usepackage{url}
\usepackage{verbatim}
\usepackage[lofdepth,lotdepth]{subfig}
\usepackage{cite}
\begin{document}
\title{2HDM with $Z_2$ symmetry in light of new LHC data}
\author{Maria Krawczyk, Dorota Soko\l owska, Bogumi\l a \'Swie\.zewska}
\address{Faculty of Physics, University of Warsaw, Ho\.{z}a 69, 00-681 Warsaw, Poland}
\ead{krawczyk@fuw.edu.pl,dsok@fuw.edu.pl, bogumila.gorczyca@fuw.edu.pl}
\date{\today}

\begin{abstract}
Properties of the $Z_2$-symmetric Two Higgs Doublet Models (2HDM) are discussed and confronted with new LHC data
for a 125 GeV Higgs particle. The particle discovered at LHC in 2012 has properties expected for it in the Standard Model (SM), with  a possible  enhancement in the $\gamma \gamma$ channel.
SM-like Higgs scenarios can be realized in the 
Two Higgs Doublet Models with $Z_2$ ($D$) symmetry  within the       
      normal Mixed Model (with scalar sector as in MSSM)
and the Inert Doublet Model (IDM),   where  a good Dark Matter (DM) candidate is present.
          Here we discuss both of the models.
\end{abstract}
\section{Introduction}
In the Two Higgs Doublet Models (2HDM) there are two doublets of  SU(2), with the weak hypercharge Y=1.   
    They give masses to $W$, $Z$ (leading at the tree-level to $\rho$ =1) and in principle also to the  photon.     
    Fermion masses are generated via Yukawa interactions, for which  various models are considered: Model I, II, III, IV, X, Y,...~\cite{Branco:2011}.
Five  scalars appear in these models, two charged   $H^+$ and  $H^-$  and  three neutral ones. If  CP is conserved there are two  CP-even $h, H$ 
and one CP-odd $A$ particle. In the model with CP violation three neutral particles  $h_1,h_2,h_3$ with undefined CP parity appear. 

\section{$D$-symmetric 2HDM}
Study of the symmetry properties of the Lagrangian as well as of the vacuum states is crucial for understanding a real content of the theory.    
Here we assume the $Z_2$ symmetry of the potential $\phi_S\to \phi_S,\,\,\, \phi_D\to -\phi_D$, which we call  below the $D$ symmetry.\footnote{In such case CP is conserved.}

In the Mixed Model both doublets have non-zero vacuum expectation values (vevs) and are involved in the mass generation. There are  five Higgs bosons and sum rules hold for the relative couplings of the neutral Higgs particles $h, H, A$:  e.g. $ (\chi^h_{VV})^2+ (\chi^H_{VV})^2+(\chi^A_{VV})^2 =1$, $V=W/Z$.  Model II is assumed for the Yukawa interactions: doublet $\phi_S$ couples 
to the down-type quarks and charged leptons, while  $\phi_D$ couples to the up-type quarks. $D$ symmetry is  spontaneously violated. In this model
SM-like scenarios are possible for both  $h$ and $H$, with  $\chi^{h/H}_{VV} = 1$.

In contrast, in the 
Inert Doublet Model (IDM), only one doublet ($\phi_S$) is involved in the mass generation and there is only one  SM-like Higgs boson $ h$. The second doublet is inert (it has vev=0) and contains four scalars. Yukawa interactions are as in Model I, so the $D$ symmetry is exact here and the neutral scalar $H$ (or $A$) may play a role of the Dark Matter (DM).

The $D$-symmetric potential has the following form:
$$
V=-\fr{1}{2}\left[m_{11}^2(\phi_S^\dagger\phi_S)\!+\! m_{22}^2(\phi_D^\dagger\phi_D)\right]+
\fr{\lambda_1}{2}(\phi_S^\dagger\phi_S)^2\!
+\!\fr{\lambda_2}{2}(\phi_D^\dagger\phi_D)^2\\[2mm]$$ 
$$+\!\lambda_3(\phi_S^\dagger\phi_S)(\phi_D^\dagger\phi_D)\!
\!+\!\lambda_4(\phi_S^\dagger\phi_D)(\phi_D^\dagger\phi_S) +\fr{\lambda_5}{2}\left[(\phi_S^\dagger\phi_D)^2\!
+\!(\phi_D^\dagger\phi_S)^2\right],$$
with all  parameters real. We take $\lambda_5<0$ without loss of generality \cite{Krawczyk:2010}. 

Such potential has various possible extrema (vacua), with the following vacuum expectation values:\footnote{With $u,v_S,v_D $ real.}
\begin{equation}
\langle\phi_S\rangle =\frac{1}{\sqrt{2}} \begin{pmatrix}0\\ v_S\end{pmatrix}\,,\qquad \langle\phi_D\rangle = \frac{1}{\sqrt{2}}
\begin{pmatrix} u \\ v_D  \end{pmatrix}. \label{dekomp_pol}
\end{equation}
Neutral vacua are realized for $u=0$. The charged vacuum with $u\not =0$ corresponds to breaking 
of U(1)$_{QED}$ symmetry and the appearance of the massive photon. The list of possible extrema, which can be realized as local  or global minima (i.e. vacua)
is given in table \ref{listvac}. The $EWs$ case corresponds to the EW symmetry, i.e. lack of
spontaneous  breaking of EW symmetry.

\begin{table}[t]
\caption{Types of extrema for a $D$-symmetric potential.
 \label{listvac}}
 \begin{center}
{\renewcommand{\arraystretch}{1.5}
\begin{tabular}{l c}
\br
type of extremum   &    condition  \\ 
\mr
EWs  ($EW\! s$) & $u = v_D = v_S = 0$ \\ 
Inert ($I_1$) & $u = v_D = 0, v_S\not =0$ \\ 
Inertlike  ($I_2$) & $u = v_S = 0, v_D\not =0$ \\ 
Mixed  ($M$) &  $u = 0, v_S \not =0, v_D \not =0 $ \\ 
Charged ( $CB$) & $u \not = 0, v_S\not =0, v_D = 0$ \\ 
\br
\end{tabular}
}
\end{center}
\end{table}

Existence of a stable vacuum requires~\cite{Ma:1978}:
\begin{displaymath}
\la>0,\quad\lb>0,\quad\lc+\sqrt{\la\lb}>0,\quad\lczp+\sqrt{\la\lb}>0\,\,\,\, (\lczp=\la+\lb+\lc), 
\end{displaymath}
so that $R=\fr{\lambda_{345}}{\sqrt{\lambda_1}\sqrt{\lambda_2}} > -1 $.
Various vacua can be realized for various values of vev, which can be  represented  in the 
phase diagram $(\mu_1,\mu_2)$, where
$$\mu_1 =\frac{m_{11}^2}{\sqrt{\la}},\quad \mu_2=\frac{m_{22}^2}{\sqrt{\lb}}.
$$

\begin{figure}[h]
\centering
\subfloat[$R>1$]{
\includegraphics[width=.3\textwidth]{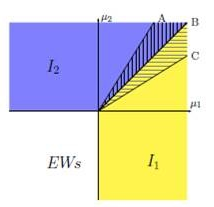}}
\subfloat[$1>R>0$]{
\includegraphics[width=.3\textwidth]{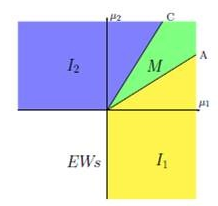}}
\subfloat[$0>R>-1$]{
\includegraphics[width=.3\textwidth]{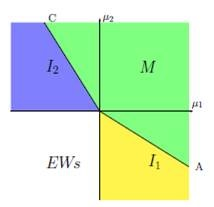}}
\caption{Phase diagrams  for $D-$symmetric potential. 
Regions of $\mu_1, \mu_2$ where various neutral extrema (minima) $EW\! s, I_1,I_2, M$ can be realized are shown. The hatched region on the left panel corresponds to the coexistence of $I_1, I_2$ minima.} \label{phasedia}
\end{figure}

There are three  regimes of parameter $R$ which correspond to very different phase patterns shown in fig. \ref{phasedia}.

In principle a model for today's Universe  could be based either on the Mixed or the Inert vacuum. 
From theory side we assume that the considered vacua are stable  and parameters of $V$ are constrained by the perturbative unitarity: $|\Lambda_i|<8\pi$~\cite{Kanemura:1993,Akeroyd:2000,Swiezewska:2012}, where $\Lambda_i$ are the eigenvalues of the high-energy scattering matrix of the scalar sector. 
This leads to the upper limits on $\lambda_{1,2}$ equal 8.38, while $\lambda_3$ and $\lambda_{345}$ are allowed to be in the regions (-6.05, 16.53) and  (-8.10,12.38), respectively.

In any case a condition for the existence of the particular vacuum has to be fulfilled. Existence of the Mixed vacuum is equivalent to having positive scalars' masses squared (condition for the minimum) while for the Inert vacuum this is not enough since  a coexistence of Inert and Inertlike minima is possible (see fig.~\ref{phasedia}). For Inert  to be the global minimum a following  condition has to be fulfilled in addition~\cite{Krawczyk:2010}:
\begin{displaymath}
\frac{m_{11}^2}{\sqrt{\la}}>\frac{m_{22}^2}{\sqrt{\lb}}.
\end{displaymath}
From  the Higgs boson mass $M_h^2=m_{11}^2=\lambda_1 v^2=(125\g)^2 $ and unitarity limit $\lb^{\textrm{max}}=8.38$ a following limit on $m_{22}^2$ arises~\cite{Swiezewska:2012}:
\begin{equation}\label{m22bound}
m_{22}^2\lesssim 9\cdot10^4\g^2.
\end{equation}

\section{Experimental constraints}
We consider both Mixed Model and IDM, taking  into account also the following experimental constraints:
\begin{description}
\item[Electroweak Precision Tests (EWPT)] Values of $S$ and $T$ parameters are demanded to lie within $2\sigma$ ellipses in the $S,T$ plane, with the following central values~\cite{Nakamura:2010}: $S=0.03\pm0.09$, $T=0.07\pm0.08$, with correlation equal to 87\%.
\item[LEP] We apply a model-independent limit: $M_{H^\pm}>70$ GeV from the direct LEP measurements. For Mixed Model 
the lower limit for mass of $H^\pm$ is 360 GeV from $b \to s \gamma$ NLO analysis \cite{Misiak:2012}. 

For the IDM  we use the LEPI and LEPII bounds on the scalar masses~\cite{Gustafsson:2009,Gustafsson:2010}:
\be
\m +M_H>M_{W},\quad \m+ M_A>M_W, \quad
M_H+M_A >M_Z
\ee
excluding region where simultaneously: $M_H< 80\g, M_A< 100\g, M_A-M_H> 8\g.$
\item[$H$ as DM candidate] In the IDM we take  $H$ as the DM candidate, $M_H<M_A,\m$. 
\end{description}

\section{Results for the Mixed Model with a 125 GeV Higgs}
Perturbative unitarity gives the following upper limits on the Higgs masses in the Mixed Model: \\
$M_h^{\textrm{max}}=$  499 GeV, $M_H^{\textrm{max}}$ and $M_{H^\pm}^{\textrm{max}}$  equal 690 GeV, $M_A^{\textrm{max}}=$ 711 GeV.
Moreover, by setting mass of $h$ equal to 125 GeV, an important limit on $\tan \beta=v_D/v_S$ is obtained, namely  0.18 $<\tan \beta<$  5.59  (independently of $\sin (\beta-\alpha)$!).
Notice, that it is possible to have  SM-like $H$,
however $h$ needs then to have a very suppressed coupling to gauge boson, see eg. \cite{Chang:2012,Drozd:2012, Pich:2013}.
 
Even when all direct  decay widths of $h$ are as in SM,  the loop decay widths $\gamma \gamma$, $Z \gamma$ for $h$  may still be modified due to $H^\pm$ contribution, or/and the negative sign of $\chi_{f \bar f h}$ coupling \cite{Ginzburg:2001}.  The $H^\pm$ loop contribution leads to a 10 \% (5\%) suppression with respect to the SM  for $\gamma \gamma$ ($Z \gamma$). The change of sign of the $tth$ coupling has a strong influence on the decay widths $\gamma \gamma$ and  $Z \gamma$ by
changing the destructive interference  between the $t$ and $W$ contributions to the loop couplings present in the SM into the constructive one.  The enhancement with respect to SM up to 2.3 (1.2) is possible for  $\Gamma_{\gamma \gamma}$ ($\Gamma_{Z\gamma}$).    For the $hgg$ coupling  change of the relative sign of the $b$ and $t$ contributions leads to a  enhancement up to 1.25.

\section{Results for IDM with a 125 GeV Higgs }
The Universe is described by the IDM if the vacuum state is given by $I_1$. IDM predicts the existence of four dark scalars $H,\,A,
H^\pm$ and the SM-like Higgs particle $h $ (we assume its mass equal to 125 GeV).
$\lambda_{345}$ is related to a triple and quartic coupling between SM-like Higgs $h$ and DM candidate $H$. $\lambda_2$ gives the quartic DM self-couplings, while $\lambda_3$ describes the Higgs particle interaction with charged scalars.

\subsection{Relic density}
IDM provides a good DM candidate in three regions of $M_H$ \cite{Cao:2007rm, Barbieri:2006dq, Gustafsson:2007pc, Dolle:2009fn, Dolle:2009ft, LopezHonorez:2006gr, Arina:2009um, Tytgat:2007cv, Honorez:2010re,LopezHonorez:2010tb,Sokolowska:2011aa,Sokolowska:2011sb}: (i) light DM particles with mass below $10 \textrm{ GeV}$, (ii) medium mass regime of $50-150 \textrm{ GeV}$ and (iii) heavy DM of mass larger than $500 \textrm{ GeV}$.  In those regions one can get the DM relic density $\Omega_{DM}h^2$ in agreement with the astrophysical data $\Omega_{DM}h^2=0.112 \pm 0.009$ \cite{PDG}. 

This estimation of $\Omega_{DM} h^2$ may be used to constrain the  $\lambda_{345}$ coupling depending on the chosen values of masses of $H$ and other scalars \cite{Dolle:2009fn,LopezHonorez:2006gr}. $\Omega_{DM} h^2$ does not limit the $\lambda_2$ parameter, although indirect constraints come from its relation to $\lambda_{345}$ parameter through the vacuum stability constraints and existence of $I_1$ vacuum \cite{Sokolowska:2011aa,Sokolowska:2011sb}.

\begin{figure}[h]
\centering
\subfloat[$M_{A} = M_{H} + 8$ GeV]{
\includegraphics[width=0.4\textwidth]{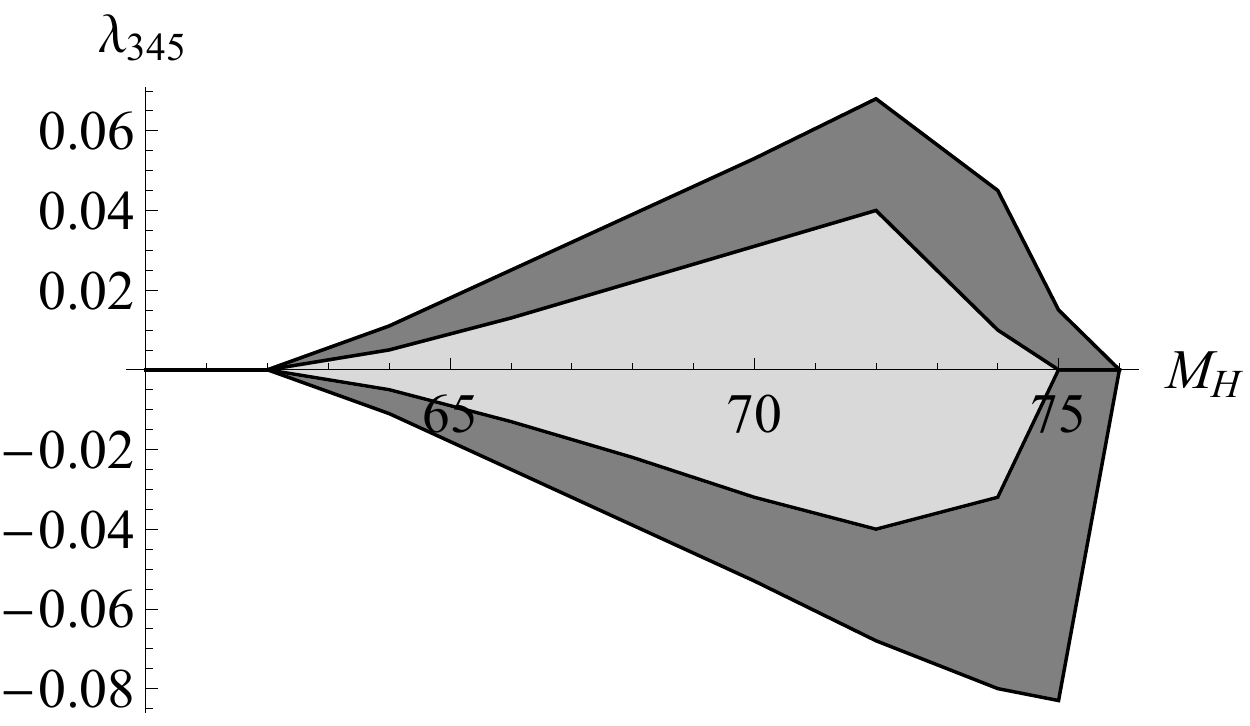}} \hspace{.5cm}
\subfloat[$M_{A} = M_{H} + 50$ GeV]{
\includegraphics[width=0.4\textwidth]{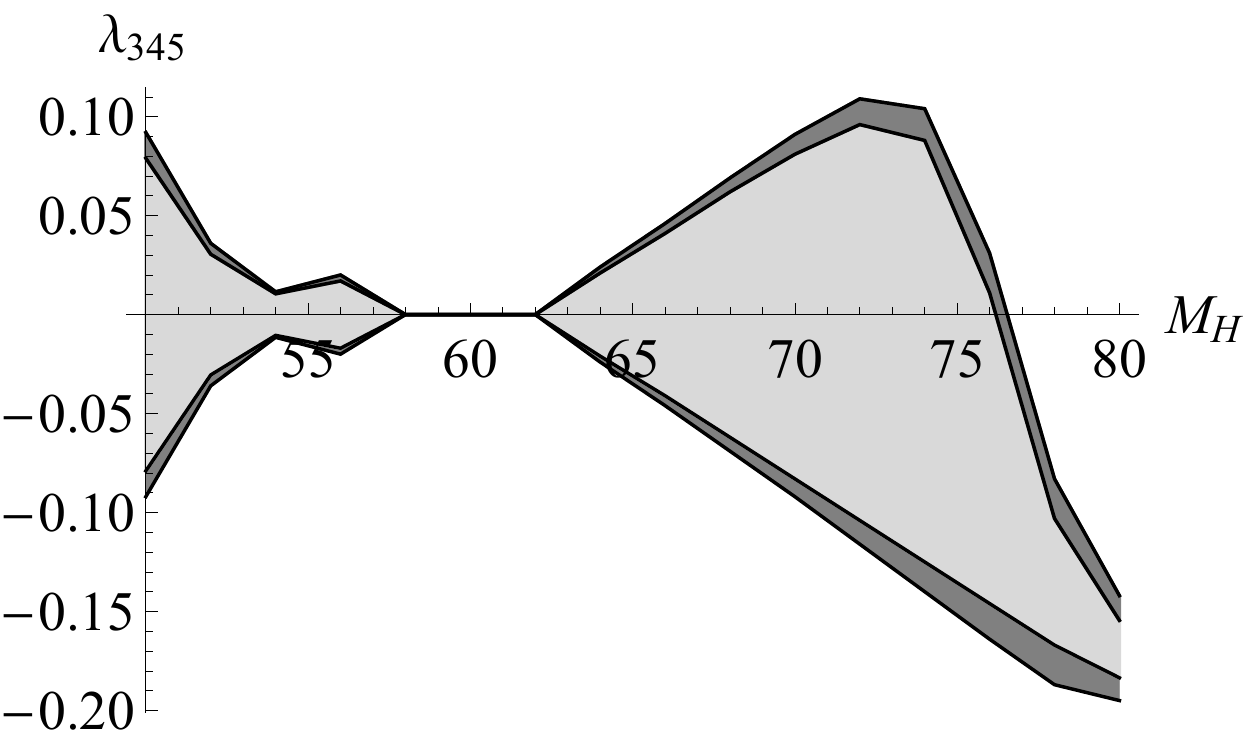}}
\caption{Limits for $(M_H, \lambda_{345})$ parameters coming from astrophysical estimations of DM relic density ($\Omega_{DM} h^2$). Dark gray: $\Omega_{DM} h^2$ in agreement with WMAP measurements, $0.1018<\Omega_{DM} h^2<0.1234$; light gray: $\Omega_{DM} h^2$ above WMAP limits (excluded); white: $\Omega_{DM} h^2$ below WMAP limits (subdominant DM). We set $M_h = 125$ GeV and $M_{H^\pm} = M_H + 50$ GeV.}\label{rDM}
\end{figure}

Limits for $M_H$ and $\lambda_{345}$ coming from $\Omega_{DM} h^2$ are presented in fig. \ref{rDM}. For masses $M_H \lesssim 72$ GeV the allowed region (dark gray) of $\lambda_{345}$ is symmetric around zero with small values of $\lambda_{345}$ excluded (light gray region) due to a nonefficient DM annhilation. As the mass increases, the region of proper relic density shifts towards the negative values of $\lambda_{345}$, which is due to opening of the annihilation channels into the gauge bosons final state.  

As it will be shown below,  in  the IDM $R_{\gamma \gamma} > 1$ is possible for $\lambda_3<0$. If we consider $H$ as a DM candidate then 
$\lambda_{345} <0$ for $R_{\gamma \gamma}>1$, meaning that it is possible to fulfill the LHC and relic density constraints for IDM.

\subsection{$\rg$}

 Here we concentrate on the 
 two-photon decay rate of the Higgs boson, which  in the IDM reads:
\begin{equation}\label{rgg}
R_{\gamma \gamma}:=\frac{\sigma(pp\to h\to \gamma\gamma)^{\textrm{IDM}}}{\sigma(pp\to h\to \gamma\gamma)^{\textrm  {SM}}}
\approx\frac{\textrm{Br}(h\to\gamma\gamma)^{\textrm {IDM}}}{\textrm{Br}(h\to\gamma\gamma)^{\textrm {SM}}}.
\end{equation}

A deviation from the value of $\rg=1$ may be caused in IDM by two factors. Firstly, the partial decay width $\Gamma(h\to \gamma\gamma)^{\textrm{IDM}}$ is modified due to the existence of the charged scalar loop~\cite{Djouadi:2005, Djouadi:2005sm, Ma:2007, Posch:2010, Arhrib:2012}:
$$
\Gamma(h\to\gamma\gamma)^{\textrm{IDM}}=\frac{G_F\alpha^2M_h^3}{128\sqrt{2}\pi^3}\left | \mathcal{M}^{\textrm{SM}}+\delta\mathcal{M}^{\textrm{IDM}}\right |^2,
$$
where $\mathcal{M}^{\textrm{SM}}$ is the contribution from the SM and $\delta\mathcal{M}^{\textrm{IDM}}=\frac{2\m^2+m_{22}^2}{2\m^2}A_0\bigg(\frac{4\m^2}{M_h^2}\bigg)$, with $2\m^2+m_{22}^2=\lc v^2$.
The charged scalar loop can interfere either constructively or destructively with the SM contribution. 
Secondly, the total decay width $\Gamma^{\textrm{IDM}}(h)$ can be increased with respect to the SM case due to the existence of the invisible decays: $h\to HH$ and $h\to AA$.

\begin{figure}[h]
\centering
\subfloat[Values of $\rg$ allowed by the theoretical and experimental constraints  as a function of the DM mass $M_H$ for $-2\cdot10^6\g^2\leqslant m_{22}^2\leqslant 9\cdot10^4\g^2$.\label{mDM}]{
\includegraphics[width=0.4\textwidth]{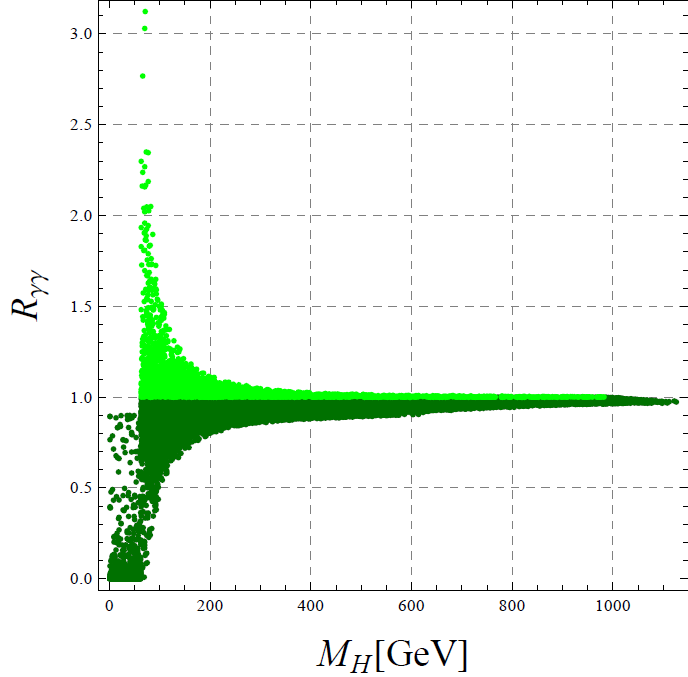}} \hspace{.5cm}
\subfloat[Region allowed by the constraints in the $(m_{22}^2,\m)$ plane. The curves correspond to the fixed values of $\rg$ (for the invisible channels closed).\label{mHp}]{
\includegraphics[width=0.4\textwidth]{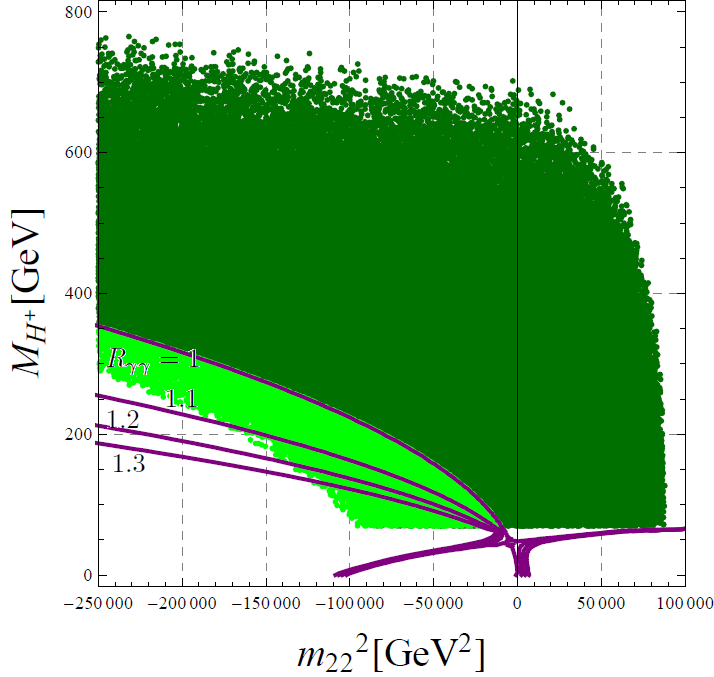}}
\caption{Results on $\rg$ for IDM, points with $\rg<1$ ($\rg>1$) are displayed in dark green/gray (light green/gray).\label{m}}
\end{figure}

Performing a random scan of the parameter space, we found the regions where $\rg $>1, with the maximal value of $\rg$  around $3.4$.
Fig.~\ref{mDM} shows values of $\rg$ allowed by the theoretical and experimental constraints as a function of $M_H$. It can be seen that enhanced values of $\rg$ are not possible for $M_H<M_h/2$. It means that if the invisible channels are open, the total decay width is so big, that it suppresses other effects.

Fig.~\ref{mHp} shows the region allowed by the theoretical and experimental constraints in the $(m_{22}^2,\m)$ plane together with the curves corresponding to constant values of $\rg$ (calculated for the case with invisible decay channels closed). It can be seen that the enhancement is possible only for constrained $m_{22}^2$ region, namely:
$$
m_{22}^2<-9.8\cdot10^3\g^2,
$$
which is equivalent to the bound $\lc<0$ (in agreement with Ref.~\cite{Arhrib:2012}). On the contrary, $\rg>1$ can be achieved for any value of $\m$. However, if bigger value of $\rg$ is demanded, then allowed values of $\m$ are constrained. For example, for $\rg>1.2$ we get the following bounds on $\m$ and $M_H$ (as $M_H<\m$): 
$$
\ba{rcccl}
62.5\g&<&M_H&<&154 \g,\\[-2pt]
70\g&<&\m&<&154 \g.\\
\ea
$$
\section{Evolution of the Universe in 2HDM through different vacua in the past }
We consider 2HDM with an explicit $D$ symmetry and assume that today the IDM describes reality. In the simplest approximation $\lambda_i$ terms in the potential are fixed and only mass terms vary with temperature:
\begin{equation}
 m_{ii}^2(T)=m_{ii}^2-c_iT^2\,\,(i=1,2),\nonumber
\end{equation}
where $ c_i=c_i (\lambda_{1-4}; g,g';\, g_t^2+g_b^2 \textrm{ for } i=1)$. Here $g, g\rq{}$ are EW gauge couplings, while  $g_t,g_b$ are the  SM Yukawa couplings)  \cite{Krawczyk:2010}.

As the Universe cools down, the quadratic coefficients vary with $T$ and the ground state of the  potential $V$ may change.
Various types of  evolution of the Universe from $EW\! s$ phase into $I_1$ phase can be realized: in one ($EW\!s \to I_1$), two ($EW\!s \to I_2 \to I_1$) or three ($EW\!s \to I_2 \to M \to I_1$) steps. In general, in $T^2$ approximation phase transitions are of the 2nd order. The only exception is the transition between two degenerate minima $I_2$ and $I_1$ in the sequence with two phase transitions. This scenario can be realized only if $R>1$. Notice, that if the Universe undergoes a series of phase transitions dark matter may appear later during the evolution, as it exists only in the Inert phase.

\begin{figure}[h]
\centering
\subfloat[$R<0$, the only sequence for $R<0$ that leads to restoration of EW symmetry in the past.\label{rR31}]{
\includegraphics[width=0.3\textwidth]{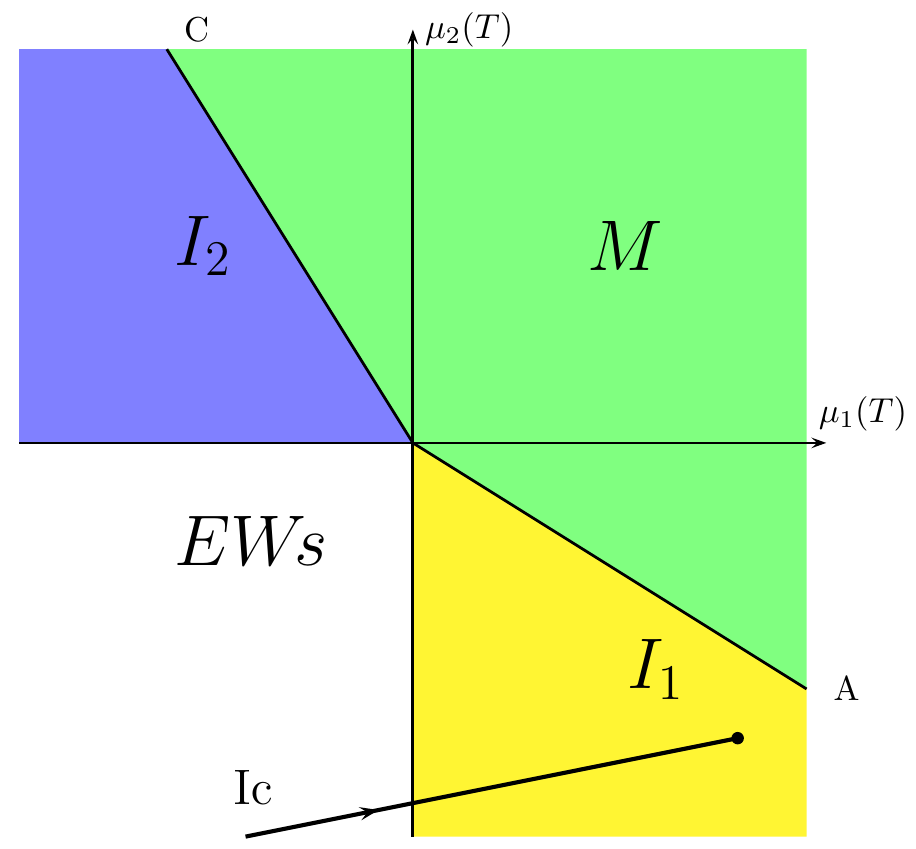}} \hspace{.2cm}
\subfloat[$R<0$, possible sequences of non-restoration of EW symmetry.\label{rR32}]{
\includegraphics[width=0.3\textwidth]{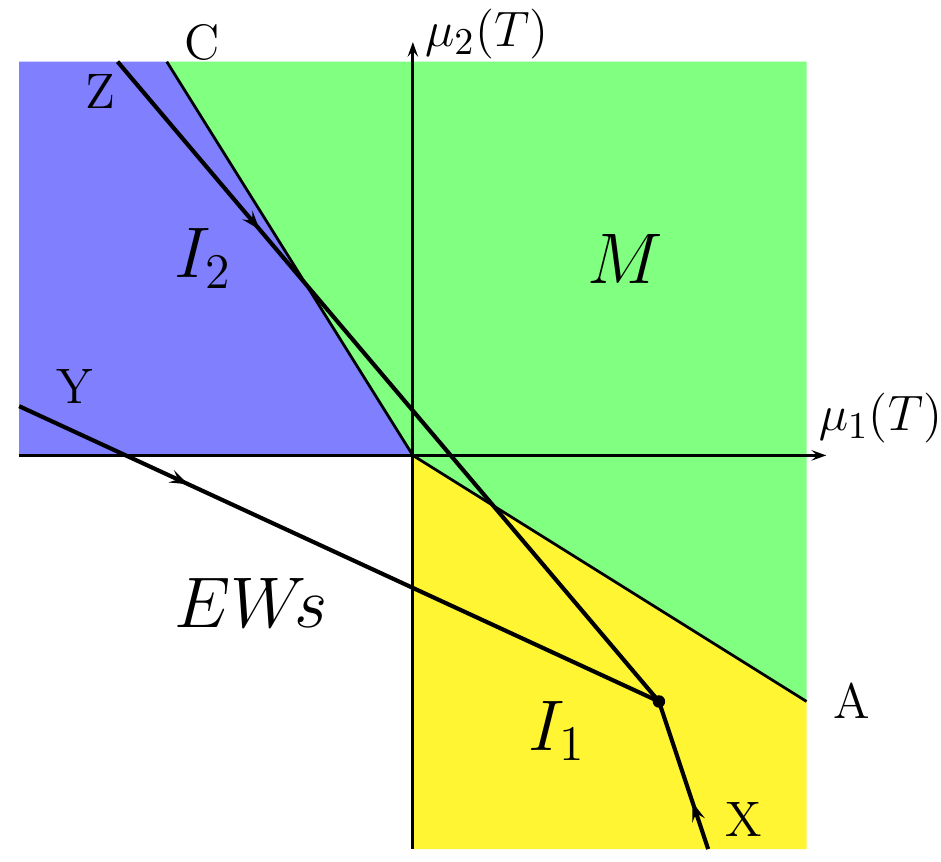}} \hspace{.2cm}
\subfloat[Transition through the charged vacuum.\label{rR33}]{
\includegraphics[width=0.3\textwidth]{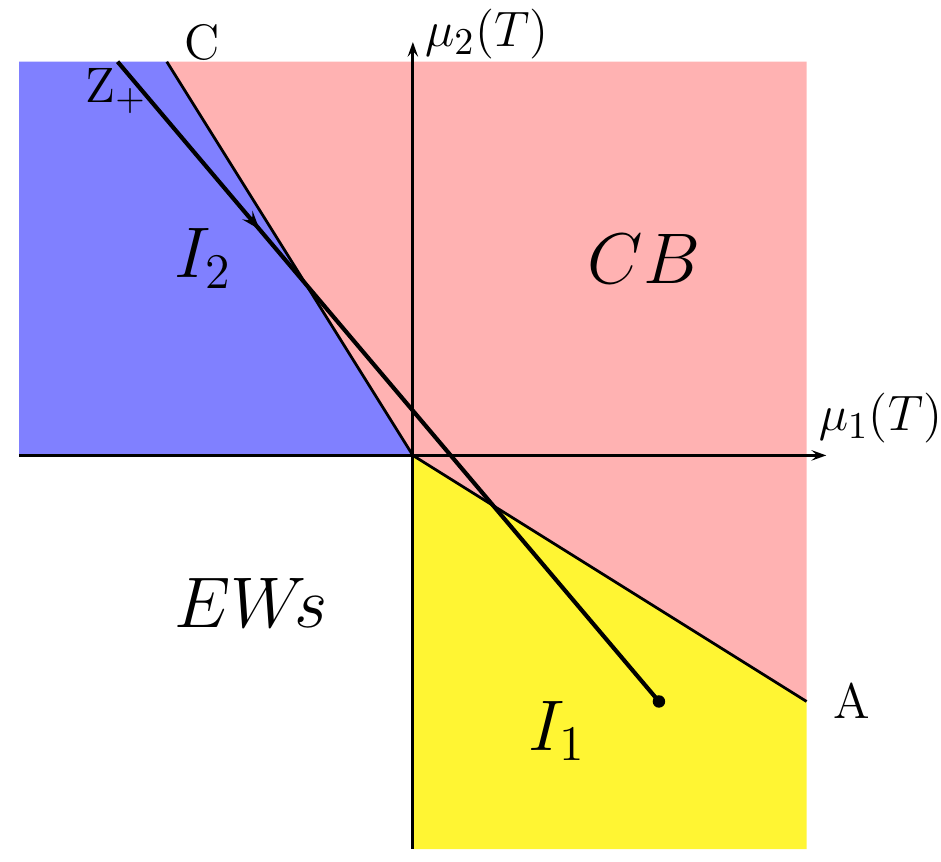}}
\caption{Possible sequences of phase transitions.}
\label{rR3}
\end{figure}

If $R<0$ there is only one type of sequence that corresponds to the restoration of EW symmetry in the past (fig. \ref{rR31}): $EW\!s \to I_1$. This sequence can be realized when $R_{\gamma \gamma} > 1$, which is suggested  by the recent LHC data. In the other scenarios for $R<0$ the initial state of the Universe is the one with broken EW symmetry (fig. \ref{rR32}). The restoration of EW symmetry may be temporary (scenario Y), in other cases the EW symmetric state never existed.

For a certain parameter range  there is a possibility of having a charged vacuum in the past (fig. \ref{rR33}). This scenario can be realized only if today we have IDM with the charged DM particle \cite{Krawczyk:2010}. 
Current model independent bounds require that  charged DM is heavier than 100 TeV \cite{Chuzhoy:2008zy}. This can be achieved without breaking the perturbative unitarity conditions with large $m_{22}^2$. However, the sequence $Z_+$ requires that $c_2<0$ and $|c_2|/c_1>|m_{22}^2|/m_{11}^2\gtrsim (10^5\div 10^6)$. This contradicts the requirement $c_1>-c_2$ based on the positivity condition \cite{Krawczyk:2010}. This means that during the evolution Universe cannot pass through  the $U(1)_{QED}$ breaking phase.

\section{Summary}
The 2HDM is an excellent laboratory of the beyond SM physics. Although the discovery of the 125 GeV Higgs particle is in agreement with the Standard Model prediction, many extensions of the SM have some build-in  SM-like scenarios. Here we consider two very different explicit $Z_2$-symmetric  extensions with two SU(2) doublets -- the very common Mixed Model and the Inert Doublet Model. Both can correspond to the SM tree-level decay modes, at the same time both may lead to the enhancement in the two-photon decay channel for the 125 GeV Higgs boson.

\ack 
MK is grateful to the organizers of the DISCRETE for an excellent conference. The work was supported by a grant  NCN OPUS 2012/05/B/ST2/03306 (2012-2016).  

\bibliographystyle{iopart-num}
\section*{References}
\bibliography{biblio}

\providecommand{\newblock}{}
\begin{thebibliography}{10}
\expandafter\ifx\csname url\endcsname\relax
  \def\url#1{{\tt #1}}\fi
\expandafter\ifx\csname urlprefix\endcsname\relax\def\urlprefix{URL }\fi
\providecommand{\eprint}[2][]{\url{#2}}

\bibitem{Branco:2011}
Branco G, Ferreira P, Lavoura L, Rebelo M, Sher M and Silva J 2012 {\em
  Phys.Rept.\/} {\bf 516} 1--102 (\textit{Preprint} \eprint{1106.0034})

\bibitem{Krawczyk:2010}
Ginzburg I, Kanishev K, Krawczyk M and Sokołowska D 2010 {\em Phys.Rev.\/} {\bf
  D82} 123533 (\textit{Preprint} \eprint{1009.4593})

\bibitem{Ma:1978}
Deshpande N~G and Ma E 1978 {\em Phys.Rev.\/} {\bf D18} 2574

\bibitem{Kanemura:1993}
Kanemura S, Kubota T and Takasugi E 1993 {\em Phys.Lett.\/} {\bf B313} 155--160
  (\textit{Preprint} \eprint{hep-ph/9303263})

\bibitem{Akeroyd:2000}
Akeroyd A~G, Arhrib A and Naimi E~M 2000 {\em Phys.Lett.\/} {\bf B490} 119--124
  (\textit{Preprint} \eprint{hep-ph/0006035})

\bibitem{Swiezewska:2012}
Swiezewska B 2012  (\textit{Preprint} \eprint{1209.5725})

\bibitem{Nakamura:2010}
Nakamura K {\em et~al.\/} (Particle Data Group) 2010 {\em J.Phys.G\/} {\bf G37}
  075021

\bibitem{Misiak:2012}
Hermann T, Misiak M and Steinhauser M 2012 {\em JHEP\/} {\bf 1211} 036
  (\textit{Preprint} \eprint{1208.2788})

\bibitem{Gustafsson:2009}
Lundstrom E, Gustafsson M and Edsjo J 2009 {\em Phys.Rev.\/} {\bf D79} 035013
  (\textit{Preprint} \eprint{0810.3924})

\bibitem{Gustafsson:2010}
Gustafsson M 2010 {\em PoS\/} {\bf CHARGED2010} 030 (\textit{Preprint}
  \eprint{1106.1719})

\bibitem{Chang:2012}
Chang J, Cheung K, Tseng P~Y and Yuan T~C 2012 {\em Int.J.Mod.Phys.\/} {\bf
  A27} 1230030 (\textit{Preprint} \eprint{1211.6823})

\bibitem{Drozd:2012}
Drozd A, Grzadkowski B, Gunion J~F and Jiang Y 2012  (\textit{Preprint}
  \eprint{1211.3580})

\bibitem{Pich:2013}
Celis A, Ilisie V and Pich A 2013  (\textit{Preprint} \eprint{1302.4022})

\bibitem{Ginzburg:2001}
Ginzburg I~F, Krawczyk M and Osland P 2001  (\textit{Preprint}
  \eprint{hep-ph/0101208})

\bibitem{Cao:2007rm}
Cao Q~H, Ma E and Rajasekaran G 2007 {\em Phys.Rev.\/} {\bf D76} 095011
  (\textit{Preprint} \eprint{0708.2939})

\bibitem{Barbieri:2006dq}
Barbieri R, Hall L~J and Rychkov V~S 2006 {\em Phys. Rev.\/} {\bf D74} 015007
  (\textit{Preprint} \eprint{hep-ph/0603188})

\bibitem{Gustafsson:2007pc}
Gustafsson M, Lundstrom E, Bergstrom L and Edsjo J 2007  (\textit{Preprint}
  \eprint{astro-ph/0703512})

\bibitem{Dolle:2009fn}
Dolle E~M and Su S 2009 {\em Phys.Rev.\/} {\bf D80} 055012 (\textit{Preprint}
  \eprint{0906.1609})

\bibitem{Dolle:2009ft}
Dolle E, Miao X, Su S and Thomas B 2010 {\em Phys.Rev.\/} {\bf D81} 035003
  (\textit{Preprint} \eprint{0909.3094})

\bibitem{LopezHonorez:2006gr}
Lopez~Honorez L, Nezri E, Oliver J~F and Tytgat M~H~G 2007 {\em JCAP\/} {\bf
  0702} 028 (\textit{Preprint} \eprint{hep-ph/0612275})

\bibitem{Arina:2009um}
Arina C, Ling F~S and Tytgat M~H 2009 {\em JCAP\/} {\bf 0910} 018
  (\textit{Preprint} \eprint{0907.0430})

\bibitem{Tytgat:2007cv}
Tytgat M~H 2008 {\em J.Phys.Conf.Ser.\/} {\bf 120} 042026 (\textit{Preprint}
  \eprint{0712.4206})

\bibitem{Honorez:2010re}
Lopez~Honorez L and Yaguna C~E 2010 {\em JHEP\/} {\bf 1009} 046
  (\textit{Preprint} \eprint{1003.3125})

\bibitem{LopezHonorez:2010tb}
Lopez~Honorez L and Yaguna C~E 2011 {\em JCAP\/} {\bf 1101} 002
  (\textit{Preprint} \eprint{1011.1411})

\bibitem{Sokolowska:2011aa}
Sokolowska D 2011 {\em Acta Phys.Polon.\/} {\bf B42} 2237 (\textit{Preprint}
  \eprint{1112.2953})

\bibitem{Sokolowska:2011sb}
Sokolowska D 2011  (\textit{Preprint} \eprint{1107.1991})

\bibitem{PDG}
Beringer~et al J  (Particle Data Group), Phys. Rev. D86, 010001 (2012)

\bibitem{Djouadi:2005}
Djouadi A 2008 {\em Phys.Rept.\/} {\bf 459} 1--241 (\textit{Preprint}
  \eprint{hep-ph/0503173})

\bibitem{Djouadi:2005sm}
Djouadi A 2008 {\em Phys.Rept.\/} {\bf 457} 1--216 (\textit{Preprint}
  \eprint{hep-ph/0503172})

\bibitem{Ma:2007}
Cao Q~H, Ma E and Rajasekaran G 2007 {\em Phys.Rev.\/} {\bf D76} 095011
  (\textit{Preprint} \eprint{0708.2939})

\bibitem{Posch:2010}
Posch P 2011 {\em Phys.Lett.\/} {\bf B696} 447--453 (\textit{Preprint}
  \eprint{1001.1759})

\bibitem{Arhrib:2012}
Arhrib A, Benbrik R and Gaur N 2012 {\em Phys.Rev.\/} {\bf D85} 095021
  (\textit{Preprint} \eprint{1201.2644})

\bibitem{Chuzhoy:2008zy}
Chuzhoy L and Kolb E~W 2009 {\em JCAP\/} {\bf 0907} 014 (\textit{Preprint}
  \eprint{0809.0436})

\end{thebibliography}
\end{document}